
\magnification 1200
\hsize 16.2 true cm             
\baselineskip = 7.600truemm        
\def\GC#1{\beta_{#1}}
\def\r {r}
\def\rr{r}
\def\lrr{\ln \rr}
\def\lrrd{\ln^2 \rr}
\def\lrrt{\ln^3 \rr}

\def\lr{\ln \r}
\def\lrd{\ln^2 \r}
\def\lrt{\ln^3 \r}

\def\LGd{\ln 2}
\def\LGdd{\ln^2 2}
\def\LGdt{\ln^3 2}
\def\LGdq{\ln^4 2}
\def\LGdc{\ln^5 2}
\def\A#1{a_{#1}}
\def\Z#1{\zeta(#1)}
\def\parn{\par\noindent}
\def\app{{\left({\alpha\over\pi}\right)}}
\def\PR{{\it Phys.Rev. }}

\def\ref#1{[#1]}
\def\oo{\infty}

\def\fig#1{{\rm fig. #1} }

\def\F{\Lambda}
\def\+{\hskip -2pt + \hskip -2pt}
\def\our{\ ({\rm our});\qquad}
\def\refk{\ ({\rm ref.[3]});}
\def\refkk{\ ({\rm ref.[3]}).}
\def\amu{{a_\mu^{(4)}}}
\def\ae{{a_e^{(4)}}}
\def\C#1{{ a_\mu^{(#1)}-a_e^{(#1)} }}
\def\CC#1{{\left(\C#1\right) }}
\def\#{{\hskip -.55 pt}}
\def\d${$ \displaystyle }

\rightline{\bf DFUB 93-6}
\rightline{15 May 1993}
\vskip 20truemm
\centerline{\bf The analytical contribution of some eighth-order graphs}
\par
\centerline{\bf containing vacuum polarization insertions }
\par
\centerline{\bf to the muon (g-2) in QED. }
\par
\vskip 20truemm
\centerline{S.Laporta}
\vskip 20truemm
\centerline{\it Dipartimento di Fisica, Universit\`a di Bologna,}
\centerline{\it and INFN, Sezione di Bologna,}
\centerline{\it Via Irnerio 46, I-40126 Bologna, Italy}
\vskip 1truemm
\centerline{\tt E-mail: laporta@bo.infn.it}
\vskip 20truemm
\centerline{\bf Abstract. }  \par
The contributions to the $g-2$ of the muon from some eighth-order
(four-loop)
graphs containing one-loop and two-loop vacuum polarization insertions
have been evaluated analytically
in QED perturbation theory,
expanding the results in the
ratio of the electron to muon mass  ${(m_e / m_\mu)}$.
The results agree with the numerical evaluations
and the asymptotic analytical results already existing
 in the literature.
\vfill\eject
Recently, the sixth-order (three-loop)
 coefficient $\C3$ of the QED perturbative
expansion of the difference between muon and electron ($g$-$2$)
$$a_\mu^{QED}-a_e^{QED}= \CC2 \app^2 +\CC3\app^3 +\CC4 \app^4 + ...
\eqno(1) $$
was calculated in closed analytical form [1][2].
The eighth-order (four-loop) coefficient $\C4$ is known only in numerical
form [3];
its more recent value is [4]: $$ \C4=127.55(41) \ . \eqno(2) $$
In this work we have calculated in  analytical form
the contributions to $\C4$ from some graphs containing insertions
on a single photon line
of one-loop and two-loop vacuum polarization subdiagrams (see fig.1);
the considered graphs are shown in figs.(2) and (3).

We found convenient to express the results expanding them in the ratio
of the electron and muon masses $(m_e/m_\mu)$;
for the sake of extensive numerical checks we have
calculated the terms containing up to $(m_e/m_\mu)^{16}$.
 Unfortunately the coefficient of each term
 becomes more and more cumbersome as the
power of $(m_e/m_\mu)$ increases, so that
we will list here the terms of the expansions
up to ($m_e/m_\mu)^2$ only.

The analytical expressions of the contributions to the muon anomaly
of the graphs shown in figs.(2) and (3),
accounting for the proper multiplicity factors, are ($r\equiv m_e/m_\mu$):

$$ \eqalign { \amu[{\rm fig.2(a)}]=
&\#-\#{4\over 27} \lrt
 \#-\#{25\over 27} \lrd
 \#-\#\left({2\over 27}\pi^2\#+\#{317\over 162}\right) \lr
 \#-\#{2\over 9} \Z3 \#-\#{25\over 162} \pi^2 \#-\#{8609\over 5832} \cr
&+
  \r \left[{101\over 1536} \pi^4 \right]
\cr
& +\r^2\left[
      {16\over 9} \lrt
     +{52\over 9} \lrd
     +\left({304\over 27}+{8\over 9} \pi^2\right) \lr
     +{136\over 35} \Z3
 +{26\over 27} \pi^2
\right.\cr
&\phantom {+r^2[} \left.
+{967\over 315}
    \right]
 +O(\r^3)\ ,}  \eqno(3) $$
$$ \eqalign { \amu[{\rm fig.2(b)}]=
&
      \left(\#\#\#\#-\#{4\over 9} \pi^2 \#+\#{119\over 27}\right) \lrd
      \#+\#\#\#\left(\#\#\#\#-{2\over 27} \pi^2 \#+\#{61\over 81}
      \right)  \lr
      \#\#\#-\#{4\over 45} \pi^4 \#+\#{13\over 27} \pi^2
      \#+\#{7627\over 1944} \cr
&      +\r^2\left[
\left(-{8\over 9} \pi^2 +{230\over 27}\right) \lr  -{4\over 3} \pi^2
 +{227\over 18}
       \right] +O(\r^4 \lrt) \ ,}  \eqno(4)$$
$$ \eqalign { \amu[{\rm fig.2(c)}] =
&
\left(\hskip -2pt
       -\#{16\over 3} \Z3 \#+\#{8\over 135} \pi^2
 \#+\#{943\over 162} \right)  \lr
     \#+\#{2\over 27} \pi^4 \#-\#{2\over 45} \Z3
    \#-\#{5383\over 4050} \pi^2
  \#+\#{57899\over 9720}
\cr
&
      +\r^2\left[  -{8\over 3} \Z3  -{26\over 105} \pi^2 +{458\over 81}
       \right] +O(\r^4 \lr) \ ,}  \eqno (5)$$
$$ \eqalign { \amu[{\rm fig.2(d)}]=
&
        {1\over 3} \lrd
       +\left(-{2\over 3} \Z3 +{5\over 4}\right)  \lr
       -{25\over 18} \Z3 +{1\over 18} \pi^2 +{509\over 432} \cr
&     +\r\left[
{101\over 72} \pi^3 \GC2
-{101\over 72} \pi^4 \LGd
 +{707\over 144} \pi^2 \Z3
+{9035\over 6912} \pi^4  -{821\over 432} \pi^3
 \right. \cr
&\phantom{+\r[} \left.
 -{5081\over 648} \pi^2
       \right]  \cr
&
      +\r^2\left[
       -{16\over 3} \lrt
       -{38\over 3} \lrd
       +\left(8 \Z3-{8\over 3} \pi^2 -22\right)  \lr
\right.\cr
&
\phantom{+\r^2[} \left.
+{766\over 2025} \pi^4
+{176\over 135} \pi^2 \LGdd
 -{1408\over 45} \A4
-{176\over 135} \LGdq
 \right. \cr
&\phantom{+\r^2[} \left.
 -{3706\over 225} \Z3 -{19\over 9} \pi^2
 -{31571\over 2700}
       \right]
        +O(\r^3) \ ,} \qquad  \eqno(6)$$
$$ \eqalign{ \amu[{\rm fig.2(e)}]=
&     \left({1\over 3} \pi^2 -{119\over 36}\right) \lr
   -{2\over 3} \pi^2 \Z3 +{119\over 18} \Z3 -{1\over 9} \pi^2
   +{473\over 432} \cr
&
      +\r^2\left[
       \left({4\over 3} \pi^2 -{115\over 9} \right)  \lr
       +{8\over 9} \pi^2 -{893\over 108}
       \right]
        +O(\r^3) \ ,} \qquad \qquad \eqno(7)$$
$$ \eqalign { \amu[{\rm fig.2(f)}]=
& \lr \left({14\over 405} \pi^4 -{128\over 9} \A4 +{16\over 27} \pi^2 \LGdd
-{16\over 27} \LGdq -{26\over 27} \Z3 +{16\over 27} \pi^2 \LGd
\right. \cr
& \left.
\qquad
+{164\over 243} \pi^2
-{673\over 81}
\right)
 -{128\over 9} \A5
 +{73\over 9} \Z5 -{98\over 405} \pi^4 \LGd
+{38\over 27} \pi^2 \Z3
 \cr
&
 -{16\over 81} \pi^2 \LGdt
 +{16\over 135} \LGdc +{22\over 405} \pi^4 -{32\over 3} \A4
 +{4\over 27} \pi^2 \LGdd
 \cr
&
-{4\over 9} \LGdq
 +{1213\over 162} \Z3
 -{8\over 3} \pi^2 \LGd
 +{4873\over 2916} \pi^2
-{33335\over 3888} \cr
&
+\r^2\left[
{7\over 135} \pi^4 -{64\over 3} \A4 +{8\over 9} \pi^2 \LGdd
 -{8\over 9} \LGdq
-{521\over 30} \Z3 -{12\over 5} \pi^2 \LGd \right. \cr
& \phantom{+\r^2[}
\left.
 +{821\over 225} \pi^2  +{56\over 15}
\right]
      +O(\r^3) \ ,}\eqno(8)$$
$$ \eqalign { \amu[{\rm fig.3(a)}]=
&      \left(
{2\over 3} \Z3 -{4\over 9} \pi^2 \LGd +{10\over 27} \pi^2
\hskip -1.7pt
 -{31\over 18} \right)  \lrd
+\lr \left(
\hskip -1.7pt
      -{11\over 162} \pi^4
   +{32\over 9} \A4
     \right.
\cr
&     \left.
      +{4\over 27} \LGdq
      +{8\over 27} \pi^2 \LGdd +{14\over 3} \Z3  -{20\over 9} \pi^2 \LGd
      +{158\over 81} \pi^2 -{115\over 18}
      \right)
\cr
&
+{32\over 9} \A5
 -{143\over 36} \Z5
 -{1\over 9} \pi^2 \Z3
 -{41\over 810} \pi^4 \LGd
-{8\over 81} \pi^2 \LGdt
 -{4\over 135} \LGdc
  \cr
&
-{119\over 1620} \pi^4
 +{80\over 9} \A4
 +{20\over 27} \pi^2 \LGdd
 +{10\over 27} \LGdq
+{133\over 18} \Z3
 -{221\over 81} \pi^2 \LGd
\cr
&
 +{1133\over 486} \pi^2
 -{8719\over 1296}
\cr
&
      -\r\left[ {2\over 45} \pi^2 \right]
 \cr
&
      +\r^2\left[
-{8\over 3} \lrd
+\left( -{28\over 3} \Z3 +{56\over 9} \pi^2 \LGd -{37\over 9} \pi^2
 +6 \right)  \lr
       \right.
\cr
&
\phantom{+\r^2[}
       \left.
+{77\over 162} \pi^4 -{224\over 9} \A4 -{56\over 27} \pi^2 \LGdd
 -{28\over 27} \LGdq
-{397\over 18} \Z3 \right.
\cr
&
\phantom{+\r^2[}
\left.
 +{257\over 27} \pi^2 \LGd -{178\over 27} \pi^2
 +{157\over 27}
      \right]
      +O(\r^3) \ ,} \eqno(9) $$
$$ \eqalign{ \amu[{\rm fig.3(b)}]=
&    \left( -{2\over 27} \pi^4 +{35\over 6} \Z3 +{16\over 9} \pi^2 \LGd
        -{62\over 81} \pi^2 -{227\over 54}
     \right) \lr \cr
&     +{20\over 27} \pi^2 \Z3
      -{409\over 2160} \pi^4 -{52\over 9} \A4
      -{35\over 54} \pi^2 \LGdd -{13\over 54} \LGdq
      -{1475\over 324} \Z3 \cr
&
      +{308\over 81} \pi^2 \LGd
      +{187\over 1458} \pi^2 -{11891\over 1944} \cr
&
      +\r^2\left[
-{14\over 135} \pi^4 -{1199\over 1080} \Z3  -{64\over 27} \pi^2 \LGd
+{9959\over 6075} \pi^2 +{18367\over 1620}
       \right]
\cr &
  +O(\r^3) \ ,} \eqno(10) $$
$$ \eqalign{ \amu[{\rm fig.3(c)}]
=
&   \left( \hskip -.5 pt
      -{3\over 4} \Z3 +{1\over 2} \pi^2 \LGd -{5\over 12} \pi^2
 +{31\over 16}\right) \lr
+{3\over 2} \zeta^2(3) -\pi^2 \Z3 \LGd
\cr
&
 +{5\over 6} \pi^2 \Z3
 +{11\over 288} \pi^4
      -2 \A4 -{1\over 6} \pi^2 \LGdd -{1\over 12} \LGdq
      -{99\over 16} \Z3
\cr
& +{25\over 24} \pi^2 \LGd
  -{133\over 144} \pi^2
  +{535\over 192}
 \cr
&
 +\r\left[-{8\over 9} \pi^2 \LGd -{13\over 36} \pi^3
 +{79\over 54} \pi^2
       \right] \cr
&
      +\r^2\left[
       6 \lrd
       +\lr\left(
       -14 \pi^2 \LGd  +21 \Z3 +{37\over 4} \pi^2 -{47\over 2}\right)
       -{77\over 72} \pi^4
\right. \cr
& \phantom{+r^2[}    \left.
       +56 \A4
       +{14\over 3} \pi^2 \LGdd
       +{7\over 3} \LGdq
       +{185\over 8} \Z3 -{39\over 4} \pi^2 \LGd
 +{57\over 8} \pi^2
\right. \cr
& \phantom{+r^2[}    \left.
 +{35\over 3}
       \right] +O(\r^3\lr) \ .}  \eqno(11) $$
Here $\zeta(p)$ is the Riemann $\zeta$-function of argument $p$,
\d$ \zeta(p) \equiv \sum_{n=1}^{\infty} {1\over n^p} \ ,$
(whose first values are $\zeta(2)=\pi^2/6$, $\zeta(3)=1.202056903...$,
$\zeta(4)=\pi^4/90$, $\zeta(5)= 1.036927755...$),
\d$ a_4 \equiv \sum_{n=1}^{\infty} {1\over {2^n n^4}}
 = 0.517479061...
 \ $,
\d$ a_5 \equiv \sum_{n=1}^{\infty} {1\over {2^n n^5}}
 = 0.508400579...
 \ $,
and $\beta_2$ is the Catalan constant
\d$\beta_2\equiv\sum_{n=0}^\oo {(-1)^n\over (2n+1)^2} = 0.915965594 ...$

We note the appearance of transcendental constants $\beta_2$ in eq.(6)
and $a_5$ in eq.(8) and eq.(9); these constants appear
for the first time in a contribution to the lepton anomaly.

Of the above expressions, eqs.(3)-(11),
only the leading terms of eq.(3) and eq.(6) were already worked out
in analytical form using renormalization group techniques [3][5].
\footnote {$(^1)$}{
We want point out that, contrarily to the assertion of Appendix A
of ref.[3],
even the leading terms of the sum
of the contributions from the graphs of fig.2(e) and fig.3(c)
can be calculated in analytical form using renormalization
group techniques. In fact, the
integral $I_1$ of eq.(A.25) of ref.[3],
evaluated numerically in that work, can be worked out in analytical
form using the contributions of sixth-order graphs
containing vacuum polarization insertions
which are known in analytical form by long time
(see eq.(2.22) of ref.[6]);
the analytical value of this integral is found to be
\d$I_1 =
 {11\over 72}\pi^4
 -{2\over 3}\pi^2 \ln^2 2
 -{1\over 3}\ln^4 2
 -8a_4
 -8\Z3
+{10\over 3}\pi^2 \ln 2
 -4\pi^2
+{641\over 36}
 $
which, inserted in eq. (A.26) of ref.[3], gives the leading
terms of the sum of our eq.(7) and eq.(11).
}.
An examination of eqs.(3)-(11)
shows that the expressions of the contributions of the graphs
containing only electron loops, eqs.(3),(6),(9) and (11),
have a linear $(m_e/m_\mu)$ term;
the numerical values of the corresponding coefficients are
respectively 6.41, -5.61, \hbox{-0.44}, -2.84.
The $(m_e/m_\mu)$ expansions of the contributions of the other graphs
(containing also muon loops) begin with the $(m_e/m_\mu)^2$ term.

We list now the numerical values of eqs.(3)-(11)
obtained
using the experimental value [7] $ {(m_\mu/m_e)}=206.768 262(30)$
and taking into account all the calculated terms of
the expansions in $(m_e/m_\mu)$,
compared with the corresponding numerical values given in ref.[3]:
$$ \eqalign{
\amu[\fig{2(a)}] &\cr
\amu[\fig{2(b)}] &\cr
\amu[\fig{2(c)}] &\cr
\amu[\fig{2(d)}] &\cr
\amu[\fig{2(e)}] &\cr
\amu[\fig{2(f)}] &\cr
\amu[\fig{3(a\+b)}]&\cr
\amu[\fig{3(c)}]&\cr
}
 \eqalign{
&= 7.2230764(8)  \cr
&= 0.49407203(3) \cr
&= 0.027988322(7)\cr
&= 7.1280084(2)  \cr
&= 0.119602460(2)\cr
&= 0.33366468(1) \cr
&=-9.3427221(5)  \cr
&=-2.77885233(5) \cr
}
 \eqalign{
&\our \cr
&\our \cr
&\our \cr
&\our \cr
&\our \cr
&\our \cr
&\our \cr
&\our \cr
}
 \eqalign{
&7.2237(13) \cr
&0.4942(2) \cr
&0.0280(1)  \cr
&7.1289(23) \cr
&0.1195(1)  \cr
&0.3337(1)  \cr
-&9.3571(40)\cr
-&2.7864(45)\cr
}
\eqalign{
& \refk \cr
& \refk \cr
& \refk \cr
& \refk \cr
& \refk \cr
& \refk \cr
& \refk \cr
& \refkk \cr
}$$
The total sum is
$$ \amu[\fig{2+3}]\  = 3.2048378(8) \qquad \our \ \  3.1845(66) \ \refkk
 \eqno(12) $$

The numerical error of our results is induced by the experimental
uncertainty of $(m_\mu/m_e)$;
in order to reach such a precision
accounting of terms up to $(m_e/m_\mu)^4$ is needed.
Note that the linear terms in $(m_e/m_\mu)$ are essential to check
the results of ref.[3] within their precision:
 as an example, the contribution of the linear term of eq.(3)
is 0.031 which is about 23 times the error
of the correspondent numerical result of ref.[3].
We found that our results are in good agreement with numerical
results of ref.[3]; only the contribution $\amu[\fig{3(a+b)}]$
shows a slight disagreement, at the level of $3.6\sigma$.
This is a remarkable cross check of results,
due to the difference of the methods followed in the two derivations.

Let us now consider the contributions
to the electron anomaly from the
graphs shown in figs.(2) and (3) with $\mu$ and $e$ leptons exchanged
\footnote {$(^2)$}{
The contributions to the electron anomaly of the graphs shown
in figs.(2) and (3) with $\mu$ leptons replaced by electrons
can be found in ref.[8].
};
we list only the leading terms of the expansions $(r\equiv m_e/m_\mu)$:

$$
 \ae[\fig{2(a)},   e\leftrightarrow\mu]    =
\rr^4\left[
 -{89\over 15015} \Z3 +{87709\over 9729720}
 \right] +O\left(\rr^6\lrr\right) \ ,
\eqno(13) $$
\vskip 3truemm
\parn\d$
 \ae[\fig{2(b)},   e\leftrightarrow\mu]    =
\rr^4\left[
 {2\over 225} \lrrd +{61\over 27000} \lrr +{5809\over 1080000}
 \right] +O\left(\rr^6\lrr^2\right)   \ ,
$
\parn
\vskip -7.600 truemm
\rightline {(14)}
\vskip 7.600 truemm
\vskip 3truemm
\parn\d$
 \ae[\fig{2(c)},   e\leftrightarrow\mu]    =
\rr^2\left[ {16\over 45} \Z3 -{203\over 486} \right]
+O\left(\rr^4\lrrt\right)  \ ,
$
\parn
\vskip -7.600 truemm
\rightline {(15)}
\vskip 7.600 truemm
\vskip 3truemm
\parn \d$ \ae[\fig{2(d)},   e\leftrightarrow\mu]    =
\rr^4\left[
 -{82\over 1215} \lrr
  -{3827\over 3742200} \pi^4
        +{712\over 10395} \A4
       -{89\over 31185} \pi^2 \LGdd
\right.$
$$ \hskip 4.7 truecm \left.
  +{89\over 31185} \LGdq
       -{756121\over 32016600} \Z3
        +{2268671641\over 31120135200}
   \right]
+O\left(\rr^6\lrr\right)     \ ,                 \eqno (16)
$$
\vskip 3truemm
\parn\d$
 \ae[\fig{2(e)},   e\leftrightarrow\mu]    =
\rr^2\left[ -{82\over 729} \pi^2 +{1681\over 1458}  \right]
+O\left(\rr^4\lrrd\right)         \ ,
$
\parn
\vskip -7.600 truemm
\rightline {(17)}
\vskip 7.600 truemm
\vskip 3truemm
\parn\d$
 \ae[\fig{2(f)},   e\leftrightarrow\mu]    =
\rr^2\left[
-{14\over 2025} \pi^4
 +{128\over 45} \A4
 -{16\over 135} \pi^2 \LGdd
 +{16\over 135} \LGdq
 +{34\over 27} \Z3
\right.$
$$ \left.
 -{344\over 1215} \pi^2
 +{424\over 405}
   \right]
+O\left(\rr^4\lrrd\right)  \ ,       \quad \
        \eqno(18)
$$
\vskip 3truemm
\parn\d$
 \ae[\fig{3(a)}, e\leftrightarrow\mu]    =
-{529\over 5832}  \rr^2  +O\left(\rr^4\lrrd\right)     \ ,
$
\parn
\vskip -7.600 truemm
\rightline {(19)}
\vskip 7.600 truemm
\vskip 3truemm
\parn\d$
 \ae[\fig{3(b)}, e\leftrightarrow\mu]    =
\rr^2\left[
       -{46\over 405} \lrrd
       +{23\over 225} \lrr
       +{2\over 225} \pi^4 +{23\over 27} \Z3 -{2\over 135} \pi^2
       -{709027\over 364500}
   \right]
 $
\vskip 3truemm
\parn
\d$ \hskip 4.1 truecm + O\left(\rr^4\lrrd\right)      \ ,
$
\parn
\vskip -7.600 truemm
\rightline {(20)}
\vskip 7.600 truemm
\vskip 3truemm
\parn\d$
 \ae[\fig{3(c)}, e\leftrightarrow\mu]    =
\rr^2\left[
 {943\over 1458} \lrr
+ {1771\over 2592} \Z3  -{41\over 729} \pi^2 -{1124\over 2187}
   \right]
+O\left(\rr^4\lrr\right)             \ ;
$
\parn
\vskip -7.600 truemm
\rightline {(21)}
\vskip 7.600 truemm
\parn
due to the smallness of the ratio $(m_e/m_\mu)$, these contributions
are almost negligible, the numerical value of the sum of
eqs.(13)-(21) being
$$ \ae[{\rm fig.}2+3]=-1.796\times 10^{-4} \ . \eqno(22) $$

Using eq.(2) and our results (12) and (22)
we can work out a new slightly different
value for $\C4$:
$$ \C4 = 127.57(41) \ . \eqno(23) $$

We sketch the method used for obtaining the analytical expressions
of eqs.(3)-(11) and eqs.(13)-(21).
Quite in general the contribution to the anomaly of the muon
from a vertex graph with vacuum polarization insertions with
electron loops can be written as

$$a_{ln}\left({m_e \over m_\mu}\right) =
  {1\over{\pi}} \int^\oo_{4m_e^2} {db\over b}
                        K_{l}\left({b\over m_\mu^2}\right)
             {\rm Im \Pi_n} \left({b\over m_e^2}\right) \ , \eqno(24)$$
where \d${\rm Im \Pi_n}\left({b \over m_e^2}\right)$
is the imaginary part of
the vacuum polarization insertion in $n$th order
and
\d$K_{l}\left({b \over m_\mu^2}\right)$ is the anomaly contribution
from
some set of $l$th-order vertex graphs in which a photon line has been given
a mass $b$.
Both quantities are analytically known up to fourth order [6][9][10].

Once that the suitable expressions of $K$ and ${\rm Im \Pi}$
are inserted in eq.(24),
the contribution to the muon ($g$-$2$)
becomes a sum of one-dimensional integrals
containing square roots,
logarithms, dilogarithms and at worst trilogarithms of
the variable $b$.

Unlike ref.[2],
where analogous sixth-order integrals were
calculated in closed analytical form,
we found convenient
to calculate the integrals of eq.(24) expanding in ${(m_e/m_\mu)}$.
We split the integration region into two parts by introducing a
cut $\F$, such that  $m_e^2\ll\F\ll m_\mu^2$:
in the region where $b\le \F$
we find \d${b\over m_\mu^2}\ll 1 $ so that we can expand
\d$K_{l}\left({b\over m_\mu^2}\right)$
for small values of the argument,
while in the region where $b>\F$
we expand
\d${\rm Im \Pi_n}\left({b\over m_e^2}\right)$
for large \d$\left({b\over m_e^2}\right)$.
Integrals over these regions were calculated analytically
using the method described in ref.[2]
as functions of $m_e$, $m_\mu$ and $\F$;
summing up the analytical contributions
of the two regions the dependence on $\F$ drops out, as expected.
A similar method is used when the $e$ and $\mu$ leptons are
exchanged.

Finally, we checked that the direct numerical evaluations
of the integrals (24) are
in perfect agreement with our analytical expressions for the same
quantities.

\vskip 10 truemm
\parn
{\bf Acknowledgement }
\parn
The author wants to thank E.Remiddi for continuous discussions and
encouragement and M.J.Levine for kindly providing
his symbolic manipulation program ASHMEDAI [11].
\par
\vfill\eject
{\bf References}
\vskip 10 truemm

\item{[1]}  S.Laporta and E.Remiddi, {\it Phys.Lett.} {\bf B} 301 (1993) 440.
\item{[2]}  S.Laporta,
            Dipartimento  di Fisica Universit\`a di Bologna,  preprint
             DFUB  92-17 (1992),
            to be published in {\it Nuovo Cimento A} {\bf 106} (1993).

\item{[3]}  T.Kinoshita and W.B.Lindquist, \PR D {\bf 41} (1990) 593;
\par
            T.Kinoshita and W.Marciano in {\it Quantum Electrodynamics},
            ed. T.Kinoshita \par
            (World Scientific, Singapore, 1990), p.419.
\parn
\item{[4]} T.Kinoshita, Cornell University preprint CLNS 93/1186 (1993);
\par
           T.Kinoshita, Cornell University preprint CLNS 93/1187 (1993).
\item{[5]} B.Lautrup and E.de Rafael,
           {\it Nucl. Phys. B} {\bf 70} (1974) 317;
           {\it Nucl. Phys. B} {\bf 78} (1974) 576 (E).
\item{[6]} R.Barbieri and E.Remiddi,
              {\it Nucl. Phys. B} {\bf 90} (1975) 233.
\item{[7]} E. R. Cohen and B. N. Taylor, {\it Rev. Mod. Phys.}
           {\bf 59} (1987) 1121.
\item{[8]} M. Caffo, S.Turrini and E.Remiddi, {\it Phys.Rev. D}
           {\bf 30} (1984) 483.
\parn
\item{[9]} G. K\"all\'en and A.Sabry, {\it Dan. Mat. Fys. Medd.}
           {\bf 29} No.17 (1955).
\parn
\item{[10]} J.A.Mignaco and E.Remiddi,
              {\it Nuovo Cimento A}  {\bf 60} (1969) 519.
\item{[11]} M. J. Levine, U.S. AEC Report No. CAR-882-25 (1971)
                                                    (unpublished).
\parn
\vfill\eject
{\bf Figure captions}
\vskip 10 truemm
\parn
Fig.1: The irreducible fourth-order vacuum polarization subdiagram.
\vskip 6 truemm
\parn
Fig.2: Eighth-order vertex graphs obtained with insertions of
       second- and fourth-order vacuum polarization subdiagrams on
       the second-order vertex graph.
\vskip 6 truemm
\parn
Fig.3: Examples of eighth-order vertex graphs obtained with the
       insertion
       of second- and fourth-order vacuum polarization subdiagrams on
       fourth-order vertex graphs.
\parn
\bye